# Design parameters for voltage-controllable directed assembly of single nanoparticles

**Benjamin F Porter[1], Leon Abelmann[2, 3] and Harish Bhaskaran[1]**

[1] Department of Materials, University of Oxford, Parks Road, Oxford OX1 3PH

[2] Korean Institute of Science and Technology in Europe (KIST Europe GmbH), Campus E71, 66123, Germany

[3] MESA+ Research Institute, University of Twente, 7500 AE Enschede, The Netherlands

E-mail: harish.bhaskaran@materials.ox.ac.uk

**Abstract.** Techniques to reliably pick and place single nanoparticles into functional assemblies are required to incorporate exotic nanoparticles into standard electronic circuits. In this paper we explore the use of electric fields to drive and direct the assembly process, which has the advantage of being able to control the nano-assembly process at the single nanoparticle level. To achieve this, we design an electrostatic gating system, thus enabling a voltage controllable nanoparticle picking technique. Simulating this system with the nonlinear Poisson-Boltzmann equation, we can successfully characterise the parameters required for single-particle placement, the key being single particle selectivity, in effect designing a system that can achieve this controllably. We then present the optimum design parameters required for successful single nanoparticle placements at ambient temperatures, an important requirement for nanomanufacturing processes.

## 1. Introduction

In order to integrate 3D devices with the 2D nature of lithography, the placement and manipulation of nanoparticles to form complex 3D nanostructures has been a long term objective of nanoscale engineering [1]. Direct pick-and-place of building blocks has been an inspiration to scientists since it was first demonstrated using single atoms by Eigler and Schweizer [2], and research continues to push towards improving the functionality of atomic [3], molecular [4] and nanoparticle [5, 6] manipulations. However, even the ability to directly control the placement of nanoparticles at room temperature into arbitrary 3-D configurations remains an on-going objective. This is in spite of very important advances in self-assembly of nanoparticles into structures [7, 8, 9] that even recently resulted in an entire circuit [10].

The use of colloidal dispersions of electrostatically stabilised nanoparticles for designing nanoscale assemblies would be very useful for a range of 3-D device applications, such as present day back-end-of-the-line (BEOL) assembly in semiconductor manufacturing. Much recent progress in this field has been made, including results in geometry induced gating [11] as well as in probe-based particle pick up [12]. The use of a triaxial probe to generate a dielectrophoresis field that acts as a non-contact trap for dielectric nanoparticles has been recently demonstrated [12, 13]. This method is proposed to work for particles as small as 5 nm [13] and has been verified for the isolation of 100 nm polystyrene beads [12]. Although maintaining non-contact avoids potential problems with adhesive forces, the ultimate spatial positioning of targeted particles would be too inconsistent due to the randomizing effects of Brownian motion in the liquid media to consistently manufacture intended structures.

Capture of individual particles was shown to be achievable by Huang et al. [14], who utilized self-limited gating mechanisms to self-assemble single nanoparticles from colloidal gold with excellent precision. Their process used surface functionalisation to isolate single nanoparticles on a template. We propose the use of an applied voltage instead to exert a modifiable force on the nanoparticles to control their directed assembly. This force attracts a nanoparticle to a specific location where its opposing electrostatic charge will discourage additional particles from attaching to the same location. Once a particle is attached, an electrostatic force of opposite polarity would then repel the particle and drive the removal of the particle at the desired location; such techniques of detachment have been successfully demonstrated [15] and can be directly applied to the system described here. This self-limiting method would allow the consistent, accurate and controlled picking and placing of *single* nanoparticles from a gold colloid. To minimize nanoparticles from detaching from the surface on which the particle is eventually placed, a potential could be maintained until the assembly is removed from the colloid and dried. Once the statistical success of the method for this planar device has been

Porter, Abelmann and Bhaskaran

established, we foresee its integration with an AFM style probe as an effective means to pick-and-place different charged nanoscale components as a nanomanufacturing technique.

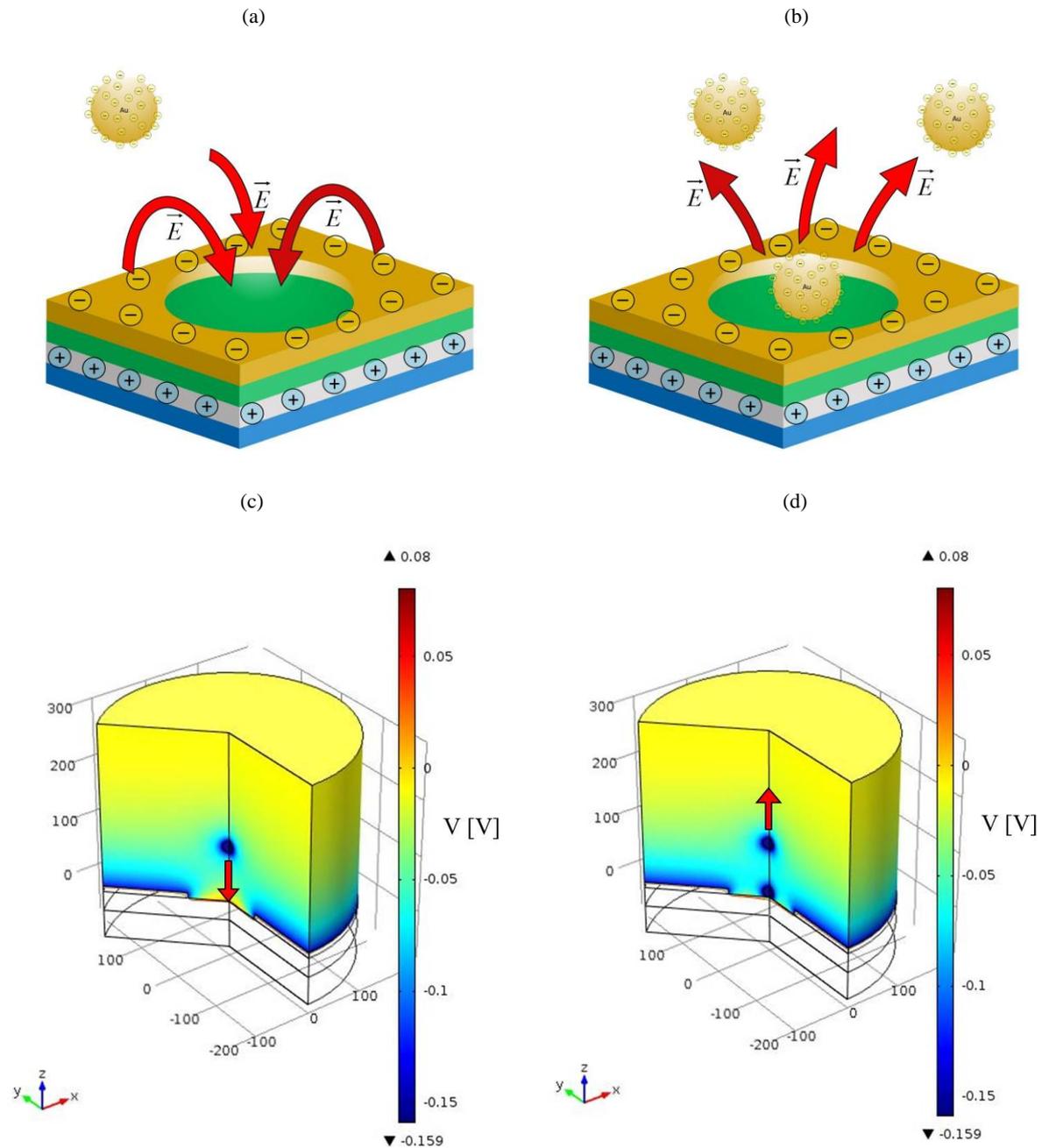

**Figure 1.** An electric bias is applied to a Ti layer (light grey), insulated from the colloid by $Si_3N_4$ (green) and patterned with Au (gold), functionalized with a self-assembled monolayer of 16-mercaptohexadecanoic acid for a negatively charged surface in solution. (a) The electric field attracts negatively charged 20 nm Au particles in the solution towards the holes (gates) in the Au layer, whilst the negative surface potential results in funnelling towards the centre of each gate. (b) Once one particle is attached to the surface of the gate, it provides a negative screening charge, with the resulting electric field preventing deposition of further nanoparticles. (c) & (d) The volume electric potential of an (c) unplugged and (d) plugged gate with an Au nanoparticle in close proximity above the gate. In these models the gate is 130 nm in diameter, the $Si_3N_4$

insulation layer is 50 nm thick, a 1 V bias is applied through the Ti underlay and the zeta potentials of the patterned Au and nanoparticles are -135 mV and -159 mV respectively. The red arrows represent the force acting on the particle above the gate.

## 2. Concept and Theory

To achieve this, we consider a structure as shown in figure 1a, whereby a dielectric is sandwiched between two conductive sheets. The top conductive layer is then patterned to form an array of circular holes (hereafter gates), where the dielectric layer is in direct contact with the solution. For the purposes of our analysis we consider the dielectric to be $Si_3N_4$ with Au gates patterned onto it. The Au top layer is functionalised with a self-assembled monolayer (SAM) of 16-mercaptohexadecanoic acid, which imparts a negative charge to the surface because of the dissociation of hydrogen ions from the hydroxyl terminus [14]. This provides an opposing potential that counters the positive electric charge created by the applied bias through the Ti underlay. Thus, the electrostatic potential profile for capturing a single negatively charged nanoparticle from a colloidal dispersion can be established (see figure 1c). It is noted that should the particle charge be positive, other appropriate SAMs can be chosen, and the treatment of our analysis below would be analogous.

The primary aim of this design is to be able to reliably pick up a single nanoparticle. There are two competing factors in choosing the gate diameter; if the diameter is too small, approaching or even being smaller than the particle diameter, the probability of the particle being attracted to the gate decreases, which reduces reliability of picking up nanoparticles. In addition, this would complicate the lithography required to pattern the gates in the first instance. On the other hand, if the diameter of the gate is too large, multiple nanoparticles would be attracted. Thus, the first step of our analysis is to consider gate diameters and the effect they play on single particle selectivity.

In order to model this system, we consider the distribution of electrical charges and the resultant forces acting on an Au particle approaching the gate when this patterned structure is placed in a colloid. The electric potential applied externally controls the electrostatic field distribution. The electric field through the structure was modeled by Poisson's equation and the distribution of ions within the solution was modeled by the nonlinear Poisson-Boltzmann equation [16]:

$$\nabla^2 \psi(\vec{r}) = -\frac{e}{\varepsilon_0 \varepsilon_r} \sum_i z_i \rho_i \exp\left(\frac{-z_i e \psi(\vec{r})}{kT}\right) \quad (1)$$

where $\psi(\vec{r})$ is the electric potential at position $\vec{r}$, $e$ is the electronic unit charge, $\varepsilon_0$ is the free space permittivity, $\varepsilon_r$ is the dielectric constant of water, $z_i$ and $\rho_i$ are the valency and bulk concentration respectively of ion species $i$, $k$ is Boltzmann's constant and $T$ is the absolute temperature. These ions are a direct result of the chemical reduction that produced the Au Sol [17], with the bulk concentrations directly related to the nanoparticle size. The force acting on the charged nanoparticles in this system can then be calculated by integrating the Maxwell Stress Tensor over the nanoparticle's surface [18]:

$$\vec{F} = \oiint \left\{ \varepsilon_0 \varepsilon_r \vec{E}\vec{E} - \frac{1}{2} \varepsilon_0 \varepsilon_r \vec{E}.\vec{E}\bar{\bar{I}} \right\} . \vec{n} dS \quad (2)$$

where $\vec{F}$ is the electric force calculated for all the following results, $\bar{\bar{I}}$ is the unit tensor, $\vec{n}$ is the normal vector to the surface and the electric field $\vec{E} = -\nabla \psi(\vec{r})$. We exclude considerations of the osmotic pressure, to focus on the influence of electrical forces. We used the finite element analysis software COMSOL to solve this model for various input parameters, outputting the electrostatic profile across the modeled geometry and the resulting force acting on a nanoparticle in the solution.

For values of zeta potentials and ionic concentrations we employ those calculated on funnelling nanoparticles by Huang et al. [14]. In order to maintain the zeta potentials required for manipulation with electric fields for controlling the nanoparticles, they should be kept in an aqueous solution of roughly neutral pH. The colloidal gold examined contained 20 nm Au nanoparticles with bulk ion concentrations of $[Na^+]$ = 4.2x10$^{-6}$M, $[C_6O_7H_7^-]$ = 9.7x10$^{-9}$M, $[C_6O_7H_6^{2-}]$ = 6.5x10$^{-7}$M, $[C_6O_7H_5^{3-}]$ = 1.0x10$^{-6}$M, $[H_3O^+]$ = 2.5x10$^{-7}$M and $[OH^-]$ = 4.0x10$^{-8}$M and the surface potentials for the self-assembled monolayer of 16-mercaptohexadecanoic acid and the Au nanoparticles were taken as -134 mV and -159 mV respectively, as previously determined [14, 19]. An example of the resulting force profiles for unplugged and plugged gates in this colloid is illustrated in figure 2, where the arrow field represents the resulting force vectors for the unplugged (red) and plugged (blue) regimes of the gate. This profile shows a clear change in the force profile once the gate is plugged, with repulsive forces replacing the attractive forces of an unplugged gate. The lateral forces also guide particles towards the centre of the gate, resulting in the central placement of the particle. Thus when comparing the different designs we have focused on the vertical forces above the centre of the gate. The following section details the results of our simulations and a discussion on the optimal parameters required for single particle placement.

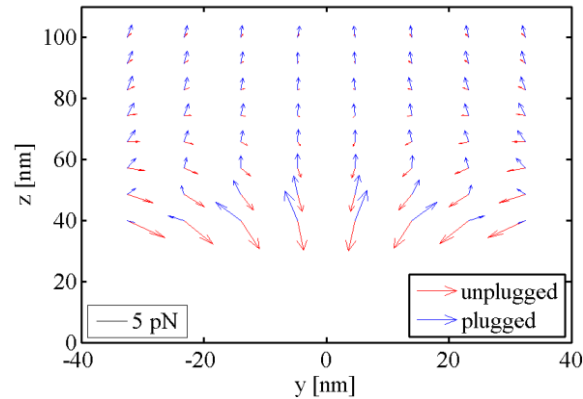

**Figure 2. Force Vectors in Liquid**: An arrow field plot of the forces acting on a nanoparticle at a range of locations above a gate for unplugged and plugged gates. The length of the vector indicates the magnitude of the force relative to the arrow in the key. The variables z and y denote the height of a particle above the gate and the radial position of the particle respectively.

Alongside these results, the equivalent forces for the static charge method established by Huang et al. [14] were also modelled (designated as "static" in the results). These serve to both validate our model as well as to offer a direct comparison for our proposed *directed* assembly system *vis-a-vis* other self-assembly approaches. As a measure of the significance of the forces exerted on particles in this colloid, we considered the characteristic Brownian force acting on these 20 nm particles as being on the order of *kT/a*: this is approximately 0.4 pN for this system, where *a* is the radius of the gold nanoparticles. To comprehensively describe the gate, we varied the diameter of the gate, the voltage applied and the thickness of the $Si_3N_4$ insulation, and the effects on the performance of the gate were analysed. (These values were typically held at 130 nm, 3 V and 50 nm respectively when other values were altered.) The principal objective of these tests was to ascertain the ideal conditions for single

Porter, Abelmann and Bhaskaran

nanoparticle capture; the largest geometrical dimensions that could reliably capture nanoparticles of much smaller diameter would enable better reproducibility in fabrication.

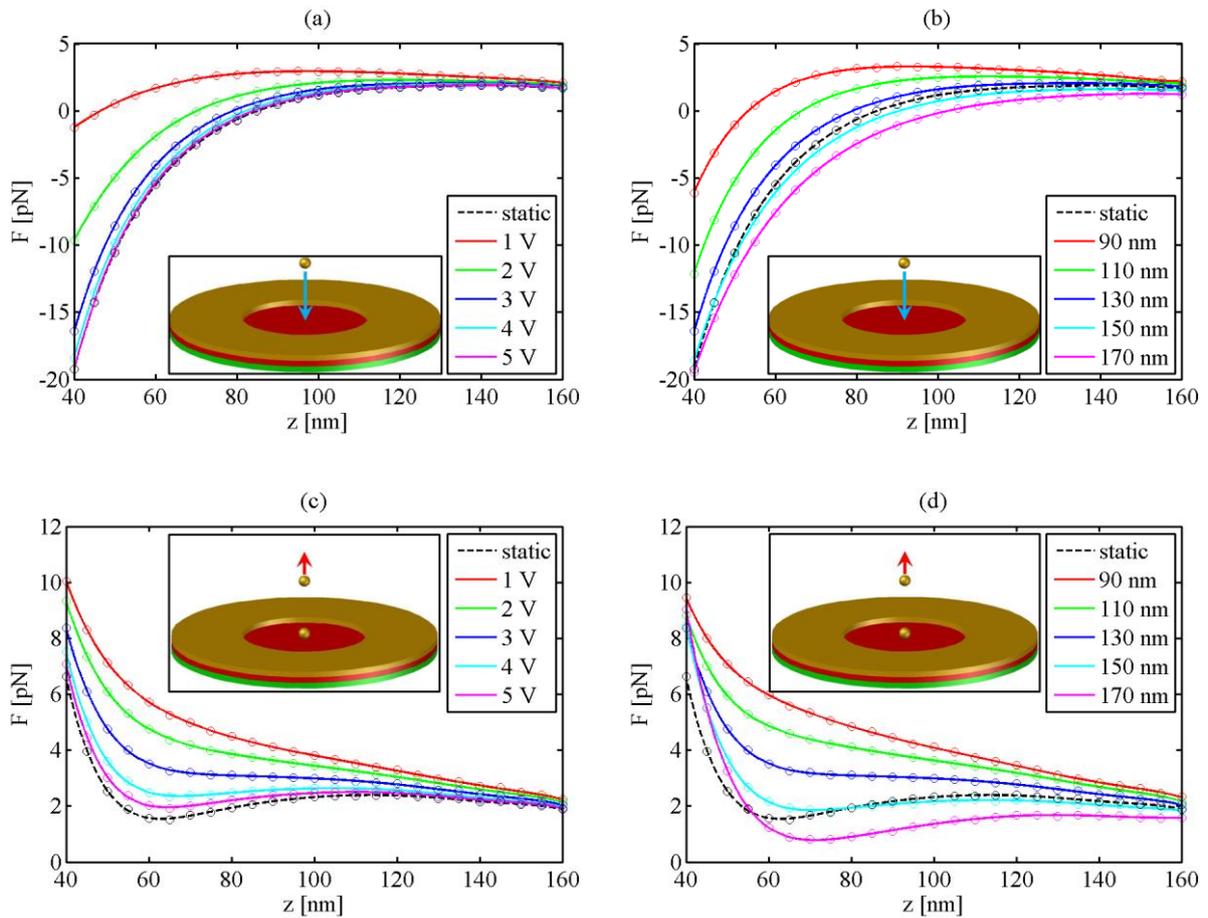

**Figure 3. Effect of Hole Diameter:** The force F acting on a nanoparticle at varying height z above the centre of a gate simulated for different voltages and gate diameters (a positive force points upwards and away from the gate and vice versa for a negative force). The target location on the gate was vacant for (a) and (b) and occupied by another nanoparticle in (c) and (d). (a) and (c) show the effects of varied voltage for a 130 nm diameter gate; (b) and (d) show the effects varied diameter for a 3 V applied bias. Static indicates the equivalent force profile for the statically charged device of Huang et al. **[14]**.

## 3. Results and Discussion

The ideal capturing technique would be able to capture one nanoparticle exactly in the center of the gate on application of a voltage; upon such a nanoparticle being captured, additional nanoparticles would be repelled away, thus resulting in reliable single particle capture. In our model described above, we capture the essential aspects of these two requirements: particle capture and subsequent particle repulsion.

From the results in Figure 3a, we can find that when no nanoparticle is captured (an unplugged gate) a 5 V bias is sufficient to create attractive forces that are strong enough to match the forces obtained by the static charge method [14]. Higher voltages for this typical geometry do not lead to substantially stronger attractive forces in the gate region. When the diameter is varied (figure 3b), we find that, as expected, larger diameters result in higher attractive forces, indicating that they are likely to attract multiple nanoparticles. Encouragingly, when a nanoparticle is captured in a gate (now a plugged gate) the resulting repulsive profile is relatively strong (figure 3c), which would prevent the approach of additional nanoparticles thus ensuring selectivity. Conversely, smaller-diameter gates are less likely to capture a nanoparticle, given the smaller magnitude of the attractive force and the lower probability of a nanoparticle being captured by the attractive force funneling vectors (figure 3d). Thus the optimum design is a balance between these two competing interests with the goal being to use the largest possible diameter that can result in single particle sensitivity.

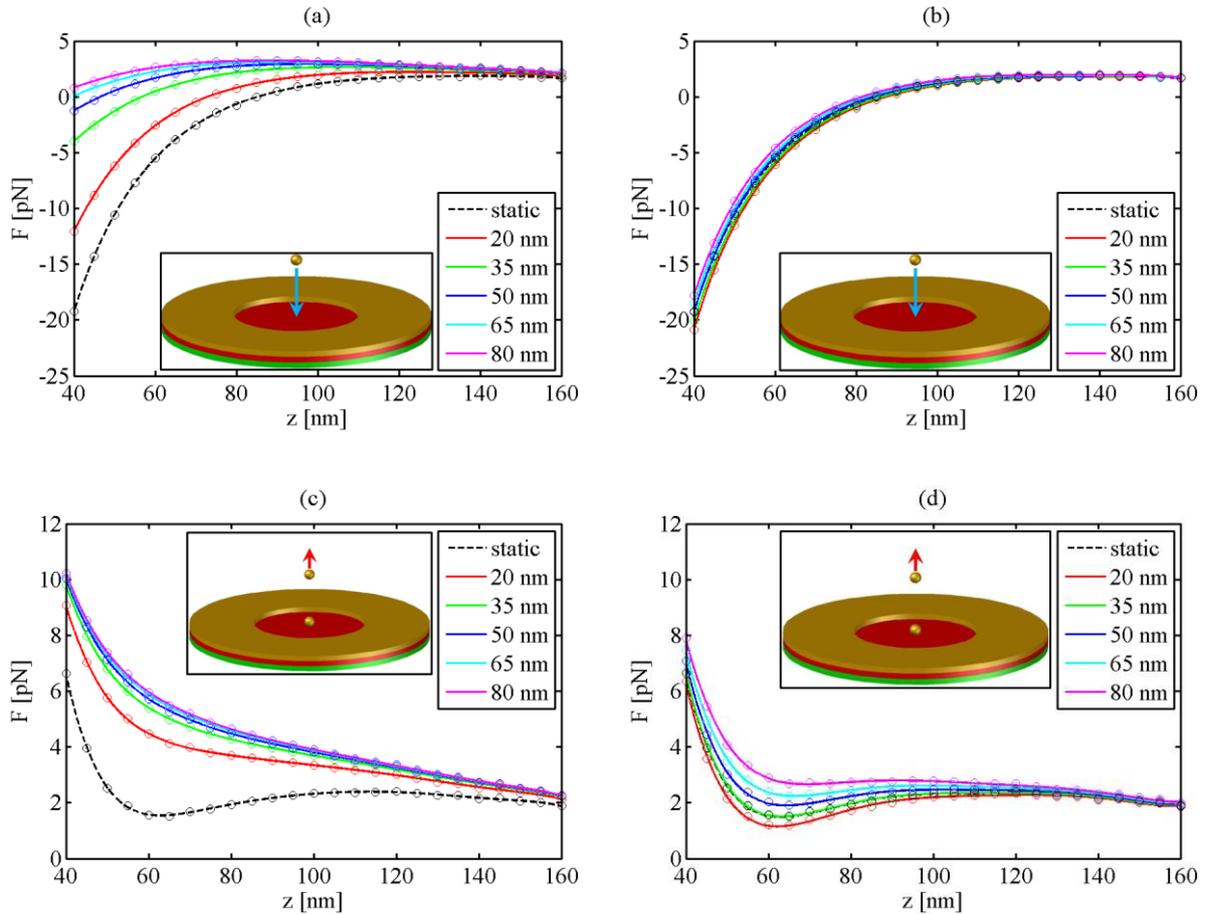

**Figure 4. Effect of Insulator Thickness.** The force F acting on a nanoparticle at varying height z above the centre of a gate simulated with varied applied voltage (a negative F is vertically down towards the gate, positive is upwards away from it). The insulation thickness was varied in all cases (see inset key), with an applied bias of 1 V in (a) and (c) and 5 V in (b) and (d). The target location on the gate was vacant for (a) and (b) and occupied by another nanoparticle in (c) and (d).

The sensitivity of the gate to changes in the voltage may also be tailored by alterations to the thickness of the insulation layer (figure 4). For instance, with a thinner $Si_3N_4$ insulation of 20 nm, a 1 V bias (figure 4a and 4c) achieves a substantially stronger attractive profile than for thicker insulations. This difference is markedly reduced when the bias is increased to 5 V (figure 4b and 4d), although the repulsive profile is still much weaker at this magnitude. From this a gate with thinner insulation and relatively low voltage (1-2 V) can be deemed the best choice. However, quality of very thick insulating layers can be highly variable, leading to imprecision in the particle placement within the gate. Optimizing this accuracy will be especially vital for the future functionality of nanomanufacturing.

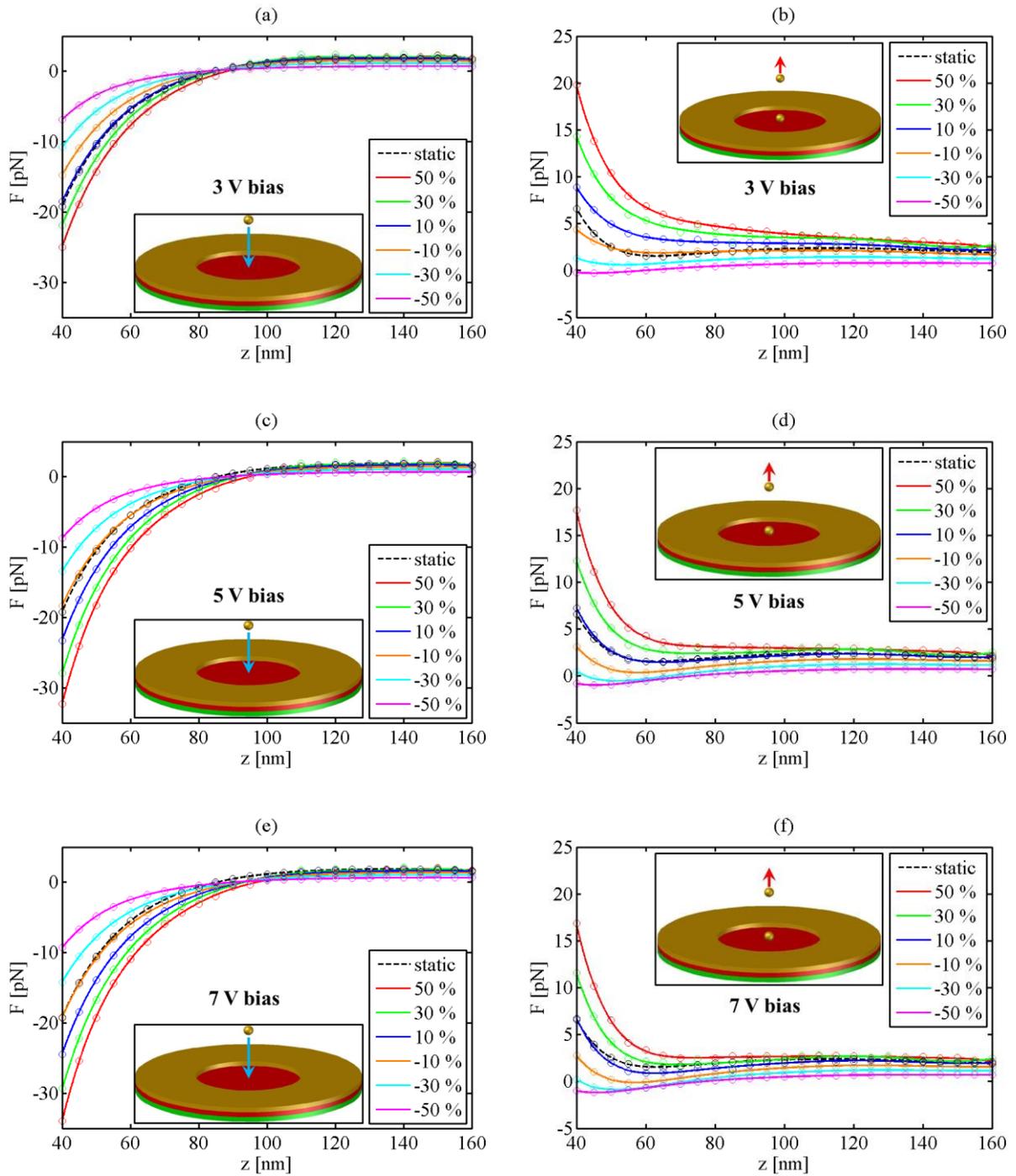

**Figure 5. Effect of particle surface potential.** The force F acting on a nanoparticle at varying height z above the centre of a gate simulated with a hypothetical 50% uncertainty in the magnitude of the Au nanoparticle's surface potential (a negative F is vertically down towards the gate, positive is upwards away from it). The applied bias was set at 3 V in (a) and (b), 5 V in (c) and (d); and 7 V in (e) and (f). The target location on the gate was vacant for (a), (c) and (e) and occupied by another nanoparticle in (b), (d) and (f).

Modifying the applied voltage is the most important feature in this design as it allows us to adjust the force exerted on the nanoparticle, which depends principally upon its surface potential. This dependency is illustrated in figure 5, where the value of the surface potential is given a hypothetical 50% uncertainty and the resulting force profiles of plugged and unplugged gates are plotted. From this we see that for nanoparticles with higher surface potentials the same placement effect appears to be more effective, with stronger attractive and repulsive forces for unplugged and plugged gates respectively. Although for lower surface potentials the attractive forces created by the applied bias are relatively weak, the repulsive forces of plugged gates are compromised by the reduced screening potential of the nanoparticle protecting the gate. For instance, for this gate a -10 % surface potential at a 3 V bias has acceptable attractive and repulsive force profiles, but the repulsion forces at -30 % are insufficient and would require a different gate diameter. Lower surface potential would likely result from a smaller particle size, so this would require reduced gate dimensions requiring lower voltage; larger particles could also be inhibited by gates that are too small. Brownian forces will have a greater impact on the motion of smaller particles, increasing the likelihood of particles breaching the weak screening potential of a particle on the gate; this suggests that a smaller gate would be needed to avoid multiple particle capture when the surface potential is radically different, but this can be estimated beforehand.

Another key design aspect is the tolerance of the particle capture accuracy to defects in the gate fabrication itself. For example, the geometry of the gate may not be the ideal circular geometry with flat surfaces as modeled. To understand how such fabrication defects would affect the nanoparticle capture, we model extreme-case scenarios of two types of defects, *viz*. a defect in the planar geometry of the gates, as well as a flatness defect on the Au layer. The results are shown in Figures 5. For example, in the first instance, we consider defects accrued by lithographic processes in the Au top layer could produce a deformed shape of the gate, as illustrated in the plan views of figure 6a. The results indicate that up to a 30% deformation of the gate geometry in either direction (relative to the intended gate radius) would have a limited effect on the equilibrium point nanoparticles will be drawn to, with inset defects causing the greater displacement. This is because the dielectric in contact with the solution is reduced, weakening the electric field that attracts the nanoparticles and making the lateral forces of the SAM relatively stronger than with offset defects. This demonstrates that the system is very tolerant to such patterning defects; for example a 10% deformation would cause a 3.3% particle placement error for the design.

The second type of defect that is not uncommon in patterning and lift-off processes is surface flatness variations. Such defects could also be caused during the Au deposition, potentially causing bulges in the layer that make the influence of the SAM electric potential on any nearby particle more prominent in that region. We model such a defect for the worst case scenario, i.e. a single defect in the metal layer near the edge of the surface. The results in figure 6b show that even if these defects were to reach unrealistic levels of 40 nm in size (which is four times the thickness of the Au layer), they would still only affect the nanoparticle placement by 3.6%. This indicates that deposition defects would have around 4 times less impact on the placement than patterning defects. This robustness is a result of the gates being essentially 3-dimensional; thus surface defects near the edges have a smaller effect in the plane.

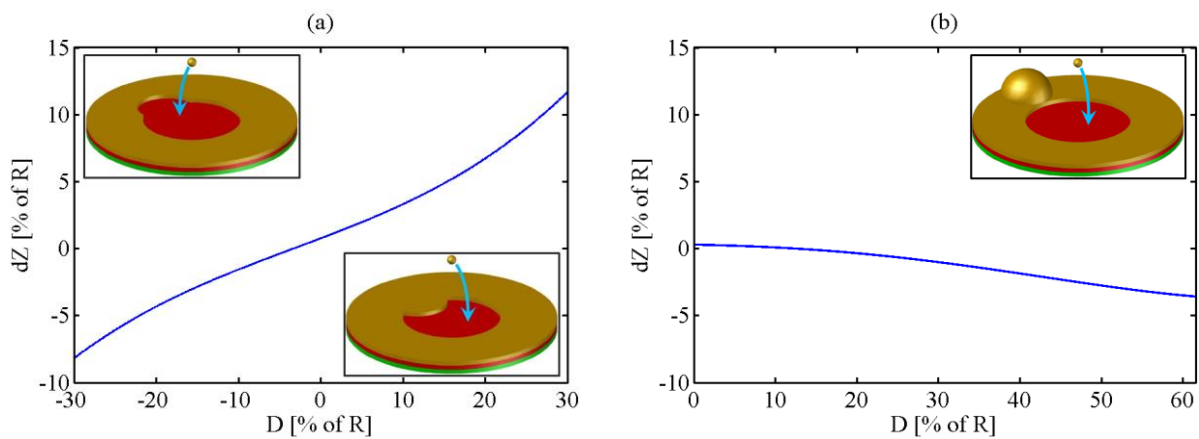

**Figure 6. Effect of Geometrical Imperfections on Particle Capture.** Testing the robustness of a gate 130 nm in diameter to fabrication defects. (a) The results of possible lithographic edge defects of size D on the placement of nanoparticles dZ, presented as percentages of the original gate radius R. (b) The results of possible Au deposition defects of size D on the placement of nanoparticles dZ as percentages of the gate radius R.

Among the exciting possibilities of our design, this work could lead on to implementing such "gates" on atomic force microscope tips, which could then be used for controlled transfer of individual nanoparticles. The limit to the maximum gate dimensions will need to be examined under experimental testing to validate the effectiveness of these designs for single nanoparticle capture. The voltage required to trap these particles is shown to be lower with a larger gate diameter, but this could compromise the central positioning of captured particles and result in multiple particle capture. Further modeling of the particle placement design could be improved by dynamic modeling, integrating the Navier-Stokes and Nernst-Planck equations into the presented Poisson-Boltzmann system [18].

## 4. Conclusions

We propose a voltage controllable, self-limiting single nanoparticle capture process in colloidal suspensions. By numerical analysis, we have optimized conditions for a novel voltage controlled pick-and-place nanoparticle assembler, investigating various configurations. We demonstrate designs that will selectively capture single nanoparticles by repelling other particles once one particle is captured. A larger gate will lead to a greater force, but increases the risk of capturing multiple particles in any one gate. Gates could be designed to isolate a range of particle sizes and surface potentials; we show that up to a 50 % increase in the surface potential of the Au nanoparticle would improve the particle selectivity. We have also shown that this design would be very resilient to fabrication defects, with lithography defects of 10% and deposition defects of 400% only effecting particle placement by 3.3% and 3.6% respectively, which indicates a very highly robust methodology. The proposed system can in essence be used for robust single nanoparticle capture with an applied voltage, and once the captured nanoparticles are placed on a target substrate, it will be able to repeat this process of capture repeatedly. Such a methodology could lead to robotic arms picking and placing nanoparticles into functional assemblies to create a directed nano-factory.

**Acknowledgements**

This work was supported by the University of Exeter, the University of Oxford and EPSRC grant EP/J018694/1.

**References**

Porter, Abelmann and Bhaskaran

[1]  G. L. Hornyak, J. Dutta, H. F. Tibbals and A. K. Rao, Introduction to nanoscience, Hoboken, New Jersey: CRC Press, 2008.

[2]  D. M. Eigler and E. K. Schweizer, "Positioning single atoms with a scanning tunnelling microscope," *Nature,* vol. 344, no. 6266, pp. 524-526., 1990.

[3]  Y. Sugimoto, M. Abe, S. Hirayama, N. Oyabu, O. Custance and S. Morita, "Atom inlays performed at room temperature using atomic force microscopy," *Nature Materials,* vol. 4, no. 2, pp. 156-159, 2005.

[4]  S. F. Heucke, F. Baumann, G. P. Acuna, P. M. D. Severin, S. W. Stahl, M. Strackharn, I. H. Stein, P. Altpeter, P. Tinnefeld and H. E. Gaub, "Placing Individual Molecules in the Center of Nanoapertures," *Nano Letters,* 2013, at press.

[5]  S. Kim, F. Shafiei, D. Ratchford and X. Li, "Controlled AFM manipulation of small nanoparticles and assembly of hybrid nanostructures," *Nanotechnology,* vol. 22, no. 115301, 2011.

[6]  S. Kim, D. C. Ratchford and X. Li, "Atomic force microscope nanomanipulation with simultaneous visual guidance," *ACSNano,* vol. 3, no. 10, pp. 2989-2994, 2009.

[7]  K. Sakakibara, J. P. Hill and K. Ariga, "Thin-film-based nanoarchitectures for soft matter: controlled assemblies into two-dimensional worlds," *Small,* vol. 7, no. 10, pp. 1288-1308, 2011.

[8]  J. F. Galisteo-Lopez, M. Ibisate, R. Sapienza, L. S. Froufe-Perez, A. Blanco and C. Lopez, "Self-assembled photonic structures," *Advanced Materials,* vol. 23, pp. 30-69, 2011.

[9]  F. S. Kim, G. Ren and S. A. Jenekhe, "One-dimensional nanostructures of π-conjugated molecular systems: assembly, properties, and applications from photovoltaics, sensors and nanophotonics to nanoelectronics," *Chemistry of Materials,* vol. 23, no. 3, pp. 682-732, 2011.

[10] H. Park, A. Afzali, S.-J. Han, G. S. Tulevski, A. D. Franklin, J. Tersoff, J. B. Hannon and W. Haensch, "High-density integration of carbon nanotubes via chemical self-assembly," *Nature Nanotechnology,* vol. 189, pp. 787-791, 2012.

[11] M. Krishnan, N. Mojarad, P. Kukura and V. Sandoghdar, "Geometry-induced electrostatic trapping of nanometric objects in a fluid," *Nature,* vol. 467, no. 7316, pp. 692-696, 2010.

[12] K. A. Brown and R. M. Westervelt, "Triaxial AFM probes for noncontact trapping and manipulation," *Nano Letters,* vol. 11, no. 8, pp. 3197-3201, 2011.

[13] K. A. Brown and R. M. Westervelt, "Proposed triaxial atomic force microscope contact-free tweezers for nanoassembly," *Nanotechnology,* vol. 20, no. 385302, 2009.

[14] H.-W. Huang, P. Bhadrachalam, V. Ray and S. J. Koh, "Single-particle placement via self-limiting electrostatic gating," *Applied Physics Letters,* vol. 93, no. 073110, 2008.

[15] Y. Zheng, C. H. Lalander, T. Thai, S. Dhuey, S. Cabrini and U. Bach, "Gutenberg-style printing of self-assembled nanoparticle arrays: electrostatic nanoparticle immobilization and DNA-mediated transfer," *Angewandte Chemie,* vol. 50, no. 19, pp. 4398-4402, 2011.

[16] J. N. Israelachvili, Intermolecular and Surface Forces, 3rd ed., Elsevier, 2011.

[17] G. Frens, "Controlled nucleation for the regulation of the particle size in monodisperse gold suspensions," *Nature,* vol. 241, pp. 20-22, 1973.

[18] J. H. Masliyah and S. Bhattacharjee, Electrokinetic and Colloid Transport Phenomena, New Jersey: John Wiley & Sons, Inc., 2006.

[19] L.-C. Ma, R. Subramanian, H.-W. Huang, V. Ray, C.-U. Kim and S. J. Koh, "Electrostatic funneling for precise nanoparticle placement: a route to wafer-scale integration," *Nano Letters,* vol. 7, no. 2, pp. 439-445, 2007.